\documentclass{ws-procs975x65}

\usepackage{graphicx,amsmath,amssymb,amsfonts}

\begin{document}

\title{\uppercase{Observable circles-in-the-sky in flat universes}}

\vspace{-1mm}
\author{BRUNO MOTA}
\address{Universidade Federal do Rio de Janeiro, \\
NACO - CCS - Av. Brigadeiro Trompowski s/n°, \ 21941-590 Rio de Janeiro -- RJ, Brazil}

\vspace{-3mm}
\author{MARCELO J. REBOU\c{C}AS}
\address{Centro Brasileiro de Pesquisas F\'{\i}sicas\\
Rua Dr.\ Xavier Sigaud 150, \  22290-180 Rio de Janeiro --
RJ, Brazil}

\vspace{-3mm}
\author{REZA TAVAKOL}
\address{Astronomy Unit, School of Mathematical Sciences, \\
Queen Mary, University of London, \  Mile End Road, London E1 4NS,
UK}

\vspace{-3mm}
\begin{abstract}
An important, and potentially detectable, signature of a
non-trivial topology for the universe is the
presence of so called circles-in-the-sky
in the cosmic microwave background (CMB).
Recent searches, confined to antipodal and nearly
antipodal circles, have however failed to detect any.
This outcome, coupled with recent theoretical results
concerning the detectability of very nearly flat
universes, is sufficient to exclude a detectable non-trivial
cosmic topology for most observers in the inflationary
limit ($0<\mid\Omega_{\text{tot}}-1\mid \lesssim10^{-5}$).
In a recent paper we have studied the consequences of these
searches for circles if the Universe turns out to be exactly
flat ($\Omega_{\text{tot}}=1$) as is often assumed.
More specifically, we have derived the maximum angles of deviation possible
from antipodicity of pairs of matching circles associated
with the shortest closed geodesic for all multiply-connected flat
orientable $3$-manifolds. These upper bounds on the deviation from
antipodicity demonstrate that in a flat universe for some classes of topology
there remains
a substantial fraction of observers for whom the deviation from
antipodicity of the matching circles is considerably
larger than zero, which implies that the searches for circles-in-the-sky
undertaken so far are not enough to exclude the possibility of
a detectable non-trivial flat topology.
Here we briefly review these results and discuss their 
consequences in the search for circles-in-the-sky in
a flat universes.
\end{abstract}

\bodymatter

\vspace{-4mm}
\section{Introduction}\label{intro}

The overall global shape of the Universe remains among the fundamental
open questions
in cosmology. The main theoretical difficulty is that, while
general relativity (as well any other local geometrical theory
of gravity) can be used to constrain the space-time  geometry,
it leaves its spatial topology undetermined (see, e.g.,
the reviews Refs.~\refcite{CosmTopReviews}).
The spatial topology of the universe, however, can in principle be
inferred by making systematic searches using the high resolution
cosmic microwave background (CMB)
data.\cite{CSS1998,Cornish-et-al-03,Key-et-al-07,Aurich-et-al-06}.
Moreover, the combination of CMB with other observational data also provides
strong support for near or exact flatness of the universe.

In standard cosmology the Universe is modeled  by
a $4$--manifold $\mathcal{M} = \mathbb{R}\times M_3$
endowed with the spatially homogeneous and isotropic
Friedmann-Lema\^{\i}tre-Robertson-Walker metric.
Furthermore, the spatial sections $M$ are often assumed
to be the simply-connected $3$--manifolds:
Euclidean $\mathbb{E}^3$, spherical $\mathbb{S}^3$, or
hyperbolic space $\mathbb{H}^3$.
However, these choices are not unique.
In fact, depending upon the spatial curvature of the three section,
there are an infinity of classes of topologically inequivalent
multiply-connected $3$--manifolds.
An important observational consequence of a non-trivial (multiply-connected)
observable spatial non-trivial topology\cite{TopDetec} of $M_3$ is the
existence of the circles-in-the-sky,\cite{CSS1998} i.e.
pairs of matching circles with the same distribution of
temperature fluctuations, identified by $\Gamma$, the so-called holonomy group.

Recently searches for antipodal and nearly antipodal (with the deviation
from antipodicity $ |\theta| \leq 10^\circ$) circles have been
undertaken, with negative results.\cite{Cornish-et-al-03,Key-et-al-07}.
In Ref.~\refcite{Mota-etal04} we have shown that a detectable spatial
topology of very nearly flat universe can be `locally'
approximated by either a slab space  ($\mathbb{R}^{2} \times  
\mathbb{S}^{1}$) or chimney space  ($\mathbb{R} \times \mathbb{T}^{2}$) 
irrespective of its global topology.
These results allow an upper bound to be placed on the
deviation from antipodicity in the inflationary limit,
which again turns out to be less than $10^\circ$  for
majority of observers.\cite{Mota-etal-08}
The combination of these results would (if confirmed)
be in principle sufficient to exclude detectable
manifolds with non-trivial (non-flat) topology for the overwhelming
majority of observers in a very nearly flat universe.\cite{Mota-etal-08}.

In a recent paper\cite{Mota-et-al-10} we have studied the consequences
of these searches for circles if the spatial section of the Universe turns
out to be exactly flat with an orientable topology.
This possibility is important not only because a flat universe is
compatible with observations, but also because it is
the assumption usually made in the so called concordance model of
cosmology.
Here we briefly summarize the main results of Ref.~\refcite{Mota-et-al-10}
and discuss their consequences in connection with the recent searches for
circles-in-the-sky.

\vspace{-5mm}
\section{Results and Concluding Remarks}

Recently we made a detailed study of the circles-in-the-sky and
in particular the maximum permissible deviations from antipodicity
in all orientable multiply-connected flat manifolds
(see Ref.~\refcite{Mota-et-al-10} for further details).
We recall that in addition to the simply connected flat
Euclidean space $\mathbb{R}^{3}$, there are a total of 17 multiply
connected three-dimensional flat manifolds which are quotient spaces
of the form of $\mathbb{R}^{3}/{\Gamma}$, where $\mathbb{R}^{3}$ is
the covering space.
Nine out of these 17 classes of manifolds are orientable.
These consist of the six compact manifolds
($E_i, i=1, \cdots ,6$): three-torus, half turn space, quarter turn space
third turn space, sixth turn space and Hantzsche-Wendt space,
plus three non-compact ones: the chimney space
$E_{11}$, the chimney space with half turn $E_{12}$ and the slab space $E_{16}$.
The manifolds, $E_1$, $E_{11}$ and $E_{16}$ are globally
homogeneous and hence would only give rise to antipodal pairs of
circles. The important question is then: What type of circles
would the remaining $6$ manifolds produce and, specifically,
what would be the maximum value for $\theta$, the deviation from
antipodicity for their most readily detectable circle pairs? .

We have calculated the maximum
values for $\theta$, $\theta_{max}$, for all orientable flat
three-manifolds with a non-trivial topology.
Our results are summarized in Table~\ref{table:theta-max}
(see Ref.~\refcite{Mota-et-al-10} for details).

\begin{table}
\tbl{Multiply-connected flat orientable $3$-manifolds and the
maximum deviation from antipodicity angles, $\theta_{\text{max}}$,
of pairs of matching circles-in-the-sky associated with the shortest
closed geodesic. \label{table:theta-max}}
{\begin{tabular}{@{}ccc@{}}
\toprule
Orientable Flat Manifold              &   Symbol       &  $\theta_{\text{max}}$ \\
\colrule
three-torus           &   $E_1$        &  $0^{\circ}$    \\
half turn space       &   $E_2$        &  $120^{\circ}$  \\
quarter turn space    &   $E_3$        &  $86^{\circ}$   \\
third turn space      &   $E_4$        &  $109^{\circ}$  \\
sixth turn space      &   $E_5$        &  $59^{\circ}$   \\
Hantzsche-Wendt space &   $E_6$        &  $120^{\circ}$  \\
chimney space         &  $E_{11}$      &  $0^{\circ}$    \\
chimney space with half turn&$E_{12}$  &  $120^{\circ}$ \\
slab space            &  $E_{16}$      &  $0^{\circ}$  \\
\botrule
\end{tabular}}
\label{aba:tbl1}
\end{table}

An important outcome of our results is that
the upper bounds on the deviation from antipodicity of
pairs of matching circles-in-the-sky
associated with the shortest closed geodesics of these manifolds can
far exceed the $\pm 10^\circ$ antipodicity interval used in
searches so far undertaken.\cite{Cornish-et-al-03,Key-et-al-07}
This has the immediate consequence that the searches so far
do not rule out the possibility of a detectable non-trivial flat topology.

\vspace{-4mm}
\section*{Acknowledgments}
M.J.R. thanks  FAPERJ (CNE) and CNPq for the grants
under which this work was carried out. 

\vspace{-4mm}


\end{document}